\documentclass[letterpaper, 10 pt, conference]{ieeeconf}
\usepackage{epsfig}
\usepackage{graphicx,graphics}
\usepackage{cite}
\usepackage{mathrsfs}
\usepackage{times}
\usepackage{mathptmx}
\usepackage{amssymb,amsmath}
\usepackage{amsfonts}
\usepackage{txfonts}
\usepackage{subfigure}
\usepackage{epstopdf}
\usepackage{mathrsfs}
\usepackage{times}
\usepackage{fancyhdr}
\usepackage{verbatim}
\usepackage{url}
\usepackage{epsfig}
\newtheorem{theorem}{\textbf{Theorem}}

\IEEEoverridecommandlockouts                              
\author{He Bai$^\dagger$ and S. Yusef Shafi$^\ddagger$
\thanks{$^\dagger$UtopiaCompression Corporation, Los Angeles, CA 90064.}
\thanks{$^\ddagger$Department of Electrical Engineering and Computer Sciences, University of California, Berkeley, CA 94720. }
\thanks{Emails: {\tt\small bluno.bai@gmail.com, yusef@eecs.berkeley.edu.}}%
}
\title{\large \bf Output Synchronization of Nonlinear Systems under Input Disturbances}
\begin{document}
\maketitle \thispagestyle{empty} \pagestyle{empty}
\begin{abstract}
We study synchronization of nonlinear systems that satisfy an incremental passivity property. We consider the case where the control input is subject to a class of disturbances, including constant and sinusoidal disturbances with unknown phases and magnitudes and known frequencies. We design a distributed control law that recovers the synchronization of the nonlinear systems in the presence of the disturbances. Simulation results of Goodwin oscillators illustrate the effectiveness of the control law. Finally, we highlight the connection of the proposed control law to the dynamic average consensus estimator developed in~\cite{BaiIMP10}.
\end{abstract}
\section{Introduction}
Synchronization of diffusively-coupled nonlinear systems is an active and rich research area \cite{Hale}, with applications to multi-agent systems, power systems, oscillator circuits, and physiological processes, among others. Several works in the literature study the case of static interconnections between nodes in full state models \cite{arcak11aut,Nijmeijer,weislotine2005,stan2007,russo2009,ScardoviEtAl,pecora1998} or phase variables in phase coupled oscillator models \cite{Kuramoto1,Strogatz,chopraspong2009,Dorfler}. Additionally, the adaptation of interconnection weights according to local synchronization errors between agents is attracting increasing attention. The authors of \cite{assenza2011} proposed a phase-coupled oscillator model in which local interactions were reinforced between agents with similar behavior and weakened between agents with divergent behavior, leading to enhanced local synchronization. Several recent works have considered adaptation strategies based on local synchronization errors \cite{zhou2006,delellis2009,yu2012}. Related problems for infinite-dimensional systems have been considered in \cite{demetriou2010,demetriou2013synchronization}. 

Common to much of the literature is the assumption that the agents to be synchronized are homogeneous with identical dynamics, and are furthermore not subject to disturbances. However, recent work has considered synchronization and consensus in the presence of exogenous inputs. In \cite{BaiIMP10}, the authors addressed the problem of robust dynamic average consensus (DAC), in which the use of partial model information about a broad class of time-varying inputs enabled exact tracking of the average of the inputs through the use of the internal model principle~\cite{IMP1} and the structure of the proportional-integral average consensus estimator formulated in \cite{freeman2006}. The problem of DAC is highly relevant to distributed estimation and sensor fusion \cite{olfati2005consensus,lynch2008,cortes2009,hatanaka2013}. In \cite{bai2011distributed}, the authors proposed an application of the internal model principle and the robust DAC estimator in~\cite{BaiIMP10} to distributed Kalman filtering. In~\cite{HBbook}, the internal model principle was used in connection with passivity to achieve adaptive motion coordination. The internal model principle has also been useful in establishing necessary and sufficient conditions for output regulation \cite{pavlov2008} and synchronization \cite{wieland2011,wieland2013,de2012}.  Reference \cite{burger2013internal} proposed internal model control strategies in which controllers were placed on the edges of the interconnection graph to achieve output synchronization under time-varying disturbances. Recent work has also addressed robust synchronization in cyclic feedback systems \cite{hamadeh2008} and in the presence of structured uncertainties \cite{dhawan2012}. 

In this paper, we consider synchronization of nonlinear systems that satisfy an incremental passivity property and are subject to a class of disturbance inputs, including constants and sinusoids with unknown phases and magnitudes and known frequencies. Constant and sinusoidal disturbances are common in control systems, due to biases in outputs of sensors and actuators, vibrations, etc. Building on the robust DAC estimator in \cite{BaiIMP10}, we design a distributed control law that achieves output synchronization in the presence of disturbances by defining an internal model subsystem at each node corresponding to the disturbance inputs. 

A key property of our approach is that local communication, computation and memory requirements are independent of the number of the systems in the network and the network connectivity, which is of interest in dense networks under processing and communication constraints. In contrast to the edge-based approach~\cite{burger2013internal}, which defines an internal model subsystem for each edge in the graph, our approach introduces such a subsystem only to each node, offering the advantage of a reduced number of internal states. Furthermore, it is easily extended to an adaptive setting where the interconnection strengths of the coupling graph are modified according to local synchronization errors.

 We next relate the control law we have derived to the robust DAC estimators studied in \cite{BaiIMP10}, and show that a specific choice of node dynamics and control law allows us to verify conditions in \cite[Theorem 2]{BaiIMP10}, guaranteeing that the output asymptotically tracks the average of the inputs. The present paper also provides a constructive approach to designing such a robust DAC estimator, which has not yet been addressed.

The rest of the paper is organized as follows. Section~\ref{sec:sync_wo} reviews the output synchronization of incrementally passive systems, and provides examples using Goodwin oscillators illustrating the effect of disturbances. Our main result on output synchronization under disturbances is presented Section~\ref{sec:main}. In Section~\ref{sec:example}, we illustrate the effectiveness of our control law using the example of Goodwin oscillators presented in Section~\ref{sec:sync_wo}. In Section~\ref{sec:dac}, we demonstrate that the control law lends a constructive approach to designing a robust dynamic average consensus estimator. Conclusions and future work are discussed in Section~\ref{sec:conclusion}.

\textit{Notation}: Let $1_N$ be the $N \times 1$
vector with all entries $1$. Let $I_N$ be the $N \times N$ identity matrix. The notation
$\mbox{diag}\{k_1,\cdots,k_n\}$ denotes the $n$ by $n$ diagonal matrix with
$k_i$ on the diagonal. Let the transpose of a real matrix $A$ be denoted by $A^T$.
%
%
%
\section{Output synchronization without input disturbances}\label{sec:sync_wo}
In this section, we briefly review the output synchronization results presented in~\cite{Stan07}, and provide an illustrative example using Goodwin oscillators.

Consider a group of $N$ identical Single-Input-Single-Output (SISO) nonlinear systems $\mathcal{H}_i$, $i=1,\cdots,N$,
given by
\begin{eqnarray}
\mathcal{H}_i: \quad \dot x_i&=&f(x_i)+g(x_i)u_i\label{dynamics}\\
  y_i&=&h(x_i).\label{output}
\end{eqnarray}
We assume that $\mathcal{H}_i$ satisfies the incremental
output-feedback passivity (IOFP) property, i.e., given two solutions
of $\mathcal{H}_i$, $x_{i_1}(t)$ and $x_{i_2}(t)$, whose
input-output pairs are $(u_{i_1}(t)\,, y_{i_1}(t))$ and
$(u_{i_2}(t)\,, y_{i_2}(t))$, there exists a positive semi-definite
incremental storage function $S(\delta x(t))\in\mathcal{C}^1$, with
$S(0) = 0$ such that
\begin{equation}
  \dot S(\delta x(t))\leq -\gamma(\delta y)^2+ \delta y\delta
  u\label{IO}
\end{equation}
where $\delta x=x_{i_1}-x_{i_2}$, $\delta y=y_{i_1}-y_{i_2}$ and
$\delta u=u_{i_1}-u_{i_2}$ and $\gamma\in\mathbb{R}$. When
$\gamma\geq 0$, $\mathcal{H}_i$ is incrementally passive (IP). When
$\gamma>0$, $\mathcal{H}_i$ is incrementally output-strictly passive
(IOSP). It is easy to show that for linear systems, passivity and
output-strict passivity are equivalent to IP and IOSP,
respectively.

\textit{Example 1: Goodwin oscillators.} Consider that each $\mathcal{H}_i$, $i=1,\ldots,4$, is a Goodwin
oscillator described by
\begin{equation}\label{eq:gw}
\mathcal{H}_i:\quad
\begin{array}{lcl}\dot x_{i_1}&=&-b_1x_{i_1}+(u_i-x_{i_4})\\
\dot x_{i_2}&=&-b_2x_{i_2}+b_2x_{i_1}\\
\dot x_{i_3}&=&-b_3x_{i_3}+b_3x_{i_2}\\
x_{i_4}&=&-\frac{1}{1+x_{i_3}^p}\\
y_i&=&x_{i_1}
\end{array}
\end{equation}
where $b_i>0$, $i = 1,2,3$. In~\cite{Stan07}, the given Goodwin oscillator model (see equation (13) and Theorem 1 in~\cite{Stan07}) was shown to be IOFP with \begin{equation}
\gamma=-\frac{-1 + \gamma_1\gamma_2\gamma_3\gamma_4\cos(\frac{\pi}{4})^4}{\gamma_1},
\end{equation}
in which $\gamma_j$ is the secant gain for the dynamics of $x_{i_k}$, $k=1,2,3$, and $\gamma_4$ is the maximum slope of the static nonlinearity $-\frac{1}{1+z^p}$ for $z>0$. Given~\eqref{eq:gw}, we have $\gamma_1 = \frac{1}{b}$, $\gamma_2=\frac{b_2}{b_2}  = 1$, and $\gamma_3=\frac{b_3}{b_3} = 1$.

In this example, we choose $b_k=0.5$, $k=1,2,3$, and $p=20$. Therefore, $\gamma_1=2$, $\gamma_2=\gamma_3=1$, and $\gamma_4=5$. Therefore, the Goodwin oscillator $\mathcal{H}_i$ is IOFP with $\gamma = -0.75$, which means that $\mathcal{H}_i$ possesses a shortage of incremental passivity. {\hspace*{\fill}$\square$}

The information flow between the $\mathcal{H}_i$ systems is described by a bidirectional and connected graph $\mathcal{G}$. If the bidirectional edge $(i,j)$ exists in $\mathcal{G}$, $y_i$ and $y_j$ are available to $\mathcal{H}_j$ and $\mathcal{H}_i$, respectively. We denote by $E$ the set of edges in $\mathcal{G}$. We define a weighted graph Laplacian matrix $L_p$ of $\mathcal{G}$, whose elements are given by 
\begin{equation}
(L_p)_{ij} = \left\{\begin{array}{cc}
\sum_{\forall j}{p_{ij}}& i=j\\
-p_{ij}& i\neq j,\\
\end{array}\right.\label{eqn:Lp}
\end{equation}
where $p_{ij}=p_{ji}\geq 0$, $p_{ij}>0$ only if $(i,j)\in E$. Since $\mathcal{G}$ is undirected, $L_p$ is symmetric and satisfies $1_N^TL_p=0_N$ and $L_p 1_N=0_N$. Let $\mu_2$ be the second smallest eigenvalue of $L_p$. Because $\mathcal{G}$ is connected, $\mu_2>0$.  

Theorem 2 in~\cite{Stan07} showed that the outputs of each $\mathcal{H}_i$ are asymptotically synchronized by the following control
\begin{equation}\label{control:no_dis}
u_i = - \sum_{(i,j)\in E}p_{ij}(y_i-y_j) \quad \forall i \in \{1,\ldots,n\},
\end{equation}
if solutions to the closed-loop system \eqref{dynamics}, \eqref{output}, and \eqref{control:no_dis} exist and $\mu_2>-\gamma$. Letting $u=[u_1,\cdots,u_N]^T$ and $y=[y_1,\cdots,y_N]^T$, we obtain a compact form of \eqref{control:no_dis}:
\begin{equation}
u=-L_py.
\end{equation}

\begin{figure}[htpb]
\centering
  \includegraphics[width=0.5\textwidth]{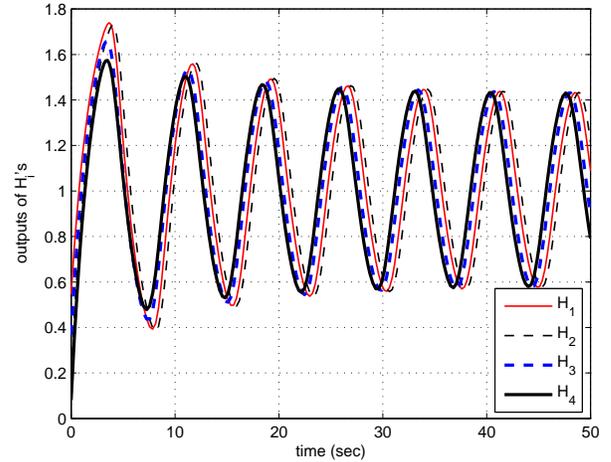}
  \caption{Because of different initial conditions, the outputs of four Goodwin oscillators are not synchronized when $u_i=0$ in~\eqref{eq:gw}.}\label{gw_nc}
\end{figure}

\textit{Example 2: Synchronization of four Goodwin oscillators.} We consider four Goodwin oscillators and use the control in \eqref{control:no_dis} to synchronize their outputs. If we choose $u_i=0$, $\forall i$, the
output of each system exhibits oscillations, as shown in Fig.~\ref{gw_nc}. Because the initial conditions of the four Goodwin models are not the same, the oscillations are out of phase.

\begin{figure}[htpb]
\centering
  \includegraphics[width=0.5\textwidth]{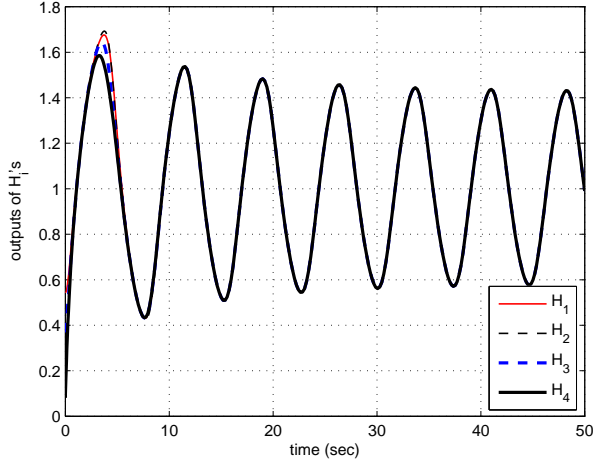}
  \caption{Adding the proportional feedback $u=-L_py$ leads to the synchronization of the four Goodwin oscillators.}\label{gw_p}
\end{figure}

We next implement the control \eqref{control:no_dis}. The graph $\mathcal{G}$ is chosen to be a cycle graph and all nonzero $p_{ij}$ in \eqref{eqn:Lp} are set to $1$. The second smallest eigenvalue of $L_p$, $\mu_2$, is $2$, satisfying $\mu_2>-\gamma$. 
Fig.~\ref{gw_p} shows that
the outputs of these four oscillators are synchronized. {\hspace*{\fill}$\square$}

\begin{figure}[htpb]
\centering
  \includegraphics[width=0.5\textwidth]{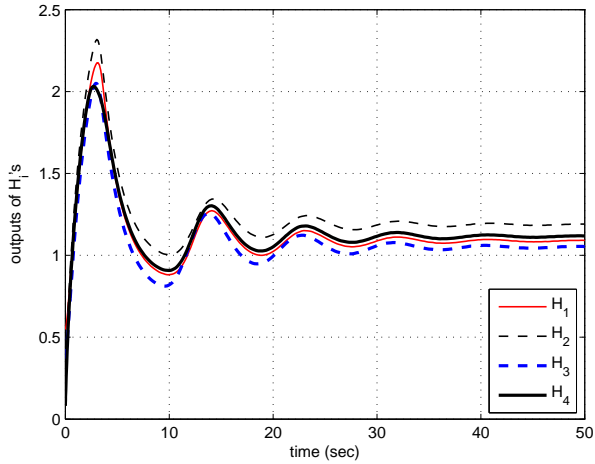}
  \caption{The nonidentical disturbance $\phi$ in the feedback $u=\phi-L_py$ destroys the synchronization of the four Goodwin oscillators.}\label{gw_p_dis}
\end{figure}

Now suppose that the input $u_i$ is subject to some constant input
disturbance $\phi_i$, $i=1,\cdots,4$. That is, $u=\phi - L_py$, where $\phi = [\phi_1,\cdots,\phi_N]^T$. The simulation result with $\phi = [0.26~0.8~0.05~0.55]^T$ is shown in Fig.~\ref{gw_p_dis}, where we observe that the outputs of the four Goodwin oscillators are not synchronized due to the nonidentical disturbances $\phi_i$.

In the next section, we present a distributed design that recovers output synchronization in the presence of a class of input disturbances, including constants and sinusoids with unknown phases and magnitudes and known frequencies. 

\section{Main result}\label{sec:main}

We consider the scenario where the input $u_i$ for each $\mathcal{H}_i$ is subject to a class of unknown disturbances $\phi_i(t)\in\mathbb{R}$, i.e., 
\begin{equation}\label{fullcontrol}
u_i = \bar u_i + \phi_i.
\end{equation}
We assume that each disturbance $\phi_i$ can be characterized by
\begin{eqnarray}
\dot\xi_i &=& A\xi_i,\quad \xi_i(0)\in\mathbb{R}^n\label{dis:state}\\
\phi_i &=& C\xi_i,\label{dis:output}
\end{eqnarray} 
in which $A\in\mathbb{R}^{n\times n}$ satisfies $A=-A^T$ and the pair $(A,C)$ is observable. Since the eigenvalues of $A$ lie on the imaginary axis, $\phi_i$ can consist of both constants and sinusoids. We assume that the matrix $A$ is available. 

Our objective is to design the control $\bar u_i$ such that the outputs of $\mathcal{H}_i$, $i=1,\cdots,N$, synchronize. We consider the following control:
\begin{equation}\label{control}
\bar u_i = - \sum_{(i,j)\in E}p_{ij}(y_i-y_j) - \sum_{(i,j)\in E}n_{ij}(\eta_i-\eta_j),
\end{equation}
where $p_{ij}$ is defined as in \eqref{eqn:Lp} and $n_{ij}=n_{ji}\geq 0$ and $n_{ij}>0$ only if $(i,j)\in E$. 
The first term in \eqref{control} is the same as \eqref{control:no_dis}. For the second term, we design $\eta_i$ to be the output of an internal model system $G_i$ given by
\begin{eqnarray}\label{IMC}
G_i: \quad \dot \zeta_i &=& A\zeta_i + B_i\sum_{(i,j)\in E}n_{ij}(y_i-y_j)\label{IMC:state}\\
\eta_i &=& B_i^T\zeta_i,\label{IMC:output}
\end{eqnarray}
where $(A,B_i^T)$ is designed to be observable and $\zeta_i(0)$, the initial condition of $\zeta_i$,  may be arbitrarily chosen.

Because $A=-A^T$ and $G_i$ is a linear system, it is straightforward to show that $G_i$ is passive and thus incrementally passive from $\sum_{(i,j)\in E}n_{ij}(y_i-y_j)$ to $\eta_i$. We will make use of the incremental passivity of $G_i$ to prove the synchronization of the outputs $y_i$ in the presence of $\phi_i$.

\begin{theorem}\label{Thm:syn} \textit{Consider the nonlinear systems $\mathcal{H}_i$ in \eqref{dynamics} and \eqref{output} satisfying \eqref{IO} with the input given in \eqref{fullcontrol}, \eqref{control}, \eqref{IMC:state} and \eqref{IMC:output}. Suppose that $\gamma+\mu_2>0$. If the solutions are bounded, then the outputs $y_i$ synchronize asymptotically:
\begin{equation}\label{eq:sync}
  \lim_{t\rightarrow \infty}|y_i(t)-\frac{1}{N}1_N^Ty(t)|=0, \quad \forall \, i \in \{1,\ldots,N\}.
\end{equation}}
\end{theorem}

\begin{figure}[htpb] 
\begin{center}
\mbox{}\setlength{\unitlength}{1.5mm}
\begin{picture}(50,37)
\put(0,0){\psfig{figure=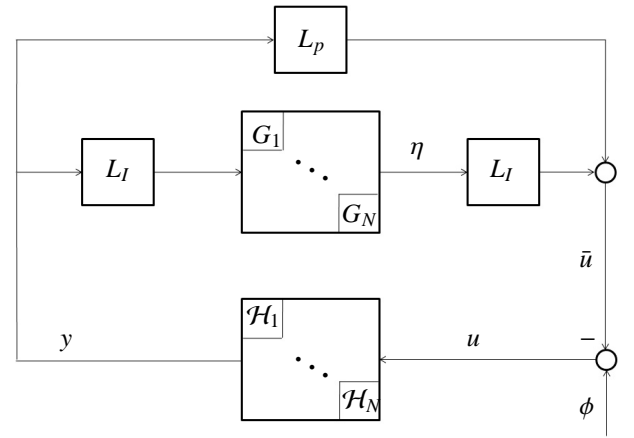,width=0.45\textwidth}}
\put(25,34.5){$L_p$} \put(4,8){$y$}\put(50,2){$
\phi$}\put(50,15){$\bar u$}\put(50,8){$-$}\put(40,8){$u$}
\put(8,23){$L_I$}\put(21,26){$G_1$}
\put(24.5,22.5){$\LARGE\ddots$}\put(29,19){$G_N$}\put(35,25){$\eta$}\put(42,23){$L_I$}
 \put(20.5,10){$\mathcal{H}_1$}\put(24.5,5){$\LARGE\ddots$}\put(29,2.5){$\mathcal{H}_N$}
  \end{picture}
\caption{The block diagram of the closed-loop system given by \eqref{dynamics}, \eqref{output}, \eqref{fullcontrol}, \eqref{control}, \eqref{IMC:state} and \eqref{IMC:output}.} \label{nonlinearstructure}
\end{center}\end{figure}

Let $\bar u_i = [\bar u_1,\cdots, \bar u_N]^T$ and $\eta = [\eta_1,\cdots,\eta_N]^T$ and define another weighted graph Laplacian $L_I$ as
\begin{equation}
(L_I)_{ij} = \left\{\begin{array}{cc}
\sum_{\forall j}{n_{ij}}& i=j\\
-n_{ij}& i\neq j,\\
\end{array}\right.\label{eqn:Li}
\end{equation}
where $n_{ij}=n_{ji}\geq 0$, $n_{ij}>0$ only if $(i,j)\in E$. Then the control in \eqref{control} can be rewritten as 
\begin{equation}\label{ubar}
\bar u = -L_py-L_I\eta.
\end{equation}
The diagram in Fig.~\ref{nonlinearstructure} shows the closed-loop system given by \eqref{dynamics}, \eqref{output}, \eqref{fullcontrol}, \eqref{control}, \eqref{IMC:state} and \eqref{IMC:output}. 

We next employ the incremental passivity property of both $\mathcal{H}_i$ and $G_i$ and the symmetry of $L_I$ to prove Theorem~\ref{Thm:syn}.
\begin{proof}
Define the orthogonal projection matrix $\Pi \in \mathbb{R}^{N \times N}$ by:
\begin{equation}\label{eq:Pi}
\Pi = I_N - \frac{1}{N}1_N1_N^T.
\end{equation} 
We first consider the storage function
\begin{equation}
  V=\frac{1}{2N}\sum_{i=1}^N\sum_{j=1}^NS(x_i-x_j),
\end{equation}
whose time derivative along (\ref{dynamics}) and (\ref{output}) is
given by
\begin{eqnarray}
  \dot V&\leq&\frac{1}{2N}\sum_{i=1}^N\sum_{j=1}^N\left[-\gamma(y_i-y_j)^2+
  (y_i-y_j)(u_i-u_j)\right]\\
  &=&-\gamma y^T \Pi y+y^T \Pi u\label{Vdot}.
\end{eqnarray}
The equality in \eqref{Vdot} follows because
\begin{eqnarray}
&&\sum_{i=1}^N\sum_{j=1}^N\left[(y_i-y_j)(u_i-u_j)\right]\\
&=&\sum_{i=1}^N\left[(y_i1_N-y)^T(u_i1_N-u)\right]\\
&=&\sum_{i=1}^N\left[Ny_iu_i+u^Ty-u_i1_N^Ty - y_i1_N^Tu\right]\\
&=&2Ny^Tu - u^T1_N1_N^Ty - y^T1_N1_N^Tu=2Ny^T\Pi u.
\end{eqnarray}
We substitute (\ref{fullcontrol}) into (\ref{Vdot}) and obtain
\begin{equation}\label{Vdot0}
\dot V\leq-\gamma  y^T\Pi y+y^T \Pi (\phi+\bar u).
\end{equation}
Noting \eqref{ubar}, we further get
\begin{eqnarray}
  \dot V&\leq&-\gamma  y^T \Pi y+y^T \Pi (\phi-L_I\eta-L_py)\nonumber\\
  &=&- y^T(\gamma \Pi+L_p)y +
  y^T( \Pi \phi-L_I\eta). \label{Vdot2}
\end{eqnarray}
We next consider auxiliary systems
\begin{eqnarray}
\dot z_i &=&Az_i,\quad z_i(0)\in\mathbb{R}^n,\quad i =1,\cdots,N,\label{aux:state}\\
\lambda_i&=&B_i^Tz_i,\label{aux:output}
\end{eqnarray} 
where the initial conditions of $z_i$, $z_i(0)$, will be chosen later, and define $\lambda = [\lambda_1,\cdots,\lambda_N]^T$. We let
\begin{equation}
\delta_i := \zeta_i-z_i,
\end{equation}
with $\delta = [\delta_1,\cdots,\delta_N]^T$. 

We claim that we can appropriately choose $z_i(0)$, $i=1,\cdots,N$, such that  
\begin{equation}\label{eq:equil}
\Pi \phi = L_I\lambda.
\end{equation}
To see this, we consider the following systems: \begin{eqnarray}
\dot{\hat\xi}_i &=& A\hat\xi_i,\quad \hat\xi_i(0)\in\mathbb{R}^n,\quad i = 1,\cdots,N\label{xi:state}\\
\hat\phi_i &=& C\hat\xi_i.\label{xi:output}
\end{eqnarray} 
We define a $(N-1)\times N$ matrix $Q$ that satisfies $Q1_N=0$, $QQ^T=I_{N-1}$ and $Q^TQ=\Pi$.
We let
\begin{equation}\label{eq:Gamma}
\Gamma = Q^T(QL_IQ^T)^{-1}Q
\end{equation}
and denote by $\Gamma_{ij}$ the element at the $i$th row and $j$th column of $\Gamma$. The inverse of $QL_IQ^T$ exists because $1_N$ spans the null spaces of $L_I$ and $Q$. Note that $\Gamma$ is the Moore-Penrose pseudoinverse of $L_I$.

We first show that choosing $\hat\xi_i(0)=\sum_{j=1}^N\Gamma_{ij}\xi_j(0)$, $\forall i$, guarantees
\begin{equation}\label{eq:phi_hat}
\Pi \phi = L_I\hat\phi
\end{equation}
where $\hat{\phi}=[\hat\phi_1,\cdots,\hat\phi_N]^T$. Note that $\hat\xi_i(0)=\sum_{j=1}^N\Gamma_{ij}\xi_j(0)$ results in $\hat{\phi}=\Gamma\phi$. Using \eqref{eq:Gamma}, we obtain 
\begin{equation}\label{eq:dphi_2}
QL_I\hat\phi = Q \phi.
\end{equation}
Pre-multiplying \eqref{eq:dphi_2} by $Q^T$ and noting $Q^TQL_I=\Pi L_I = L_I$ verify \eqref{eq:phi_hat}.

We next show that by selecting $z_i(0)$ in \eqref{aux:state} appropriately, we ensure $\lambda = \hat{\phi}$ and achieve \eqref{eq:equil} due to \eqref{eq:phi_hat}. In particular, we choose $z_i(0)=\mathcal{O}_{B_i}^{-1}\mathcal{O}_C\hat{\xi}_i(0)$, where $\mathcal{O}_{B_i}$ is the observability matrix of \eqref{aux:state}-\eqref{aux:output} and $\mathcal{O}_C$ is the observability matrix of \eqref{xi:state}-\eqref{xi:output}. Since $z_i(0)=\mathcal{O}_{B_i}^{-1}\mathcal{O}_C\hat{\xi}_i(0)$, $z_i(t)=\mathcal{O}_{B_i}^{-1}\mathcal{O}_C\hat{\xi}_i(t)$, which means $\mathcal{O}_{B_i}z_i(t)=\mathcal{O}_C\hat{\xi}_i(t)$. Noting that the first row of $\mathcal{O}_{B_i}$ and $\mathcal{O}_C$ is $B_i^T$ and $C$, respectively, we have $\lambda_i = B_i^Tz_i=C\hat{\xi}_i=\phi_i$, $\forall i$.

Having proved that \eqref{eq:equil} can be achieved by appropriately selecting $z_i(0)$ in \eqref{aux:state}, we now consider the following storage function:
\begin{equation}
  W=\frac{1}{2}\sum_{i=1}^N \delta_i^T\delta_i. 
\end{equation}
Using \eqref{IMC:state}, \eqref{IMC:output}, \eqref{aux:state} and \eqref{aux:output}, we obtain:
\begin{eqnarray}
\dot W &=& \sum_{i=1}^N\delta_i^TB_i\sum_{(i,j)\in E}n_{ij}(y_i-y_j)\\
&=&(\eta-\lambda)^TL_Iy.
\end{eqnarray}
The sum $Z = V+W$ yields
\begin{eqnarray}
\dot Z &=& \dot V + \dot W \nonumber\\&\leq& -y^T(\gamma \Pi + L_p)y+
  y^T(\Pi \phi-L_I\eta) + (\eta-\lambda)^TL_Iy\nonumber\\
  &=&-y^T(\gamma \Pi + L_p)y + y^T(\Pi \phi - L_I\lambda).
\end{eqnarray}
By choosing $z_i(0)$ in (\ref{aux:state}) such that \eqref{eq:equil} is guaranteed, we have
\begin{equation}
\dot Z\leq -y^T(\gamma \Pi + L_p)y.\label{eq:Zdot}
\end{equation} 
Noting $y^TL_p y = (Qy)^TQL_pQ^T(Qy)\geq \mu_2(Qy)^TQy$, we obtain from \eqref{eq:Zdot}
\begin{equation}\label{eq:lyapfinal}
  \dot Z\leq-(\gamma+\mu_2)y^T Q^T Qy \leq 0.
\end{equation}
By integrating both sides of (\ref{eq:lyapfinal}), we see that $Qy$ is
in $\mathcal{L}_2$. Furthermore, the boundedness of solutions implies
that $\dot{x}_i$ and thus $\dot{y}_i$ are bounded for all $i$. An
application of Barbalat's Lemma \cite{Khalil} implies that
$Qy \rightarrow 0$ as $t \rightarrow \infty$. Thus, $\Pi y \rightarrow 0$ as $t \rightarrow \infty$, which, together with \eqref{eq:Pi}, leads to \eqref{eq:sync}.
\end{proof}

We note from \eqref{IMC:state} that the differences between the outputs of the $i$th node and its neighboring nodes are first aggregated and then passed as an input to an internal model subsystem $G_i$. This node-based approach is different from the edge-based approach~\cite{burger2013internal}, where the difference between the outputs of the $i$th node and each of its neighboring nodes is directly passed to an internal model subsystem. Thus, the $i$th node needs to maintain one internal model subsystem for each of its neighboring nodes. For our node-based approach, each node maintains only one internal model subsystem in total and the dimension of $\zeta_i$ is independent of the number of the nodes in the network and the number of neighbors of the $i$th nodes. This is advantageous in dense networks under processing and communication constraints. A comparison between the performance of the node-based and edge-based approaches is currently being pursued by the authors.

In \eqref{ubar}, the two Laplacian matrices $L_p$ and $L_I$ are obtained from the same graph $\mathcal{G}$ with different weights $p_{ij}$ and $n_{ij}$. Theorem~\ref{Thm:syn} is easily extended to the case where $L_p$ and $L_I$ correspond to two different connected graphs. It is also straightforward to generalize Theorem~\ref{Thm:syn} to Multiple-Input-Multiple-Output (MIMO) systems with possibly different graphs for each output. 

Furthermore, we may incorporate adaptive updates of the weights $p_{ij}$ and $n_{ij}$ of the Laplacian matrices $L_P$ and $L_I$ according to
\begin{equation}
\begin{aligned}
\dot{p}_{ij} &= \alpha_{ij} (y_i - y_j)^2 \\
\dot{n}_{ij} &= \beta_{ij} (y_i - y_j)^2
\end{aligned}
\end{equation}
when $i$ and $j$ are neighbors in the graphs represented by $L_I$ and $L_P$, respectively, and with $\alpha_{ij}=\alpha_{ji}>0$, $\beta_{ij}=\beta_{ji}>0$. Such an update law increases the weights $p_{ij}$ and $n_{ij}$ according to the local output synchronization error between nodes $i$ and $j$, and may be implemented at each node. In order to maintain symmetry, the gains would be chosen to satisfy $\alpha_{ij}=\alpha_{ji}$ and $\beta_{ij}=\beta_{ji}$, and each pair of nodes $i$ and $j$ would update weights beginning from the same initial conditions $p_{ij}(0)=p_{ji}(0)$, $n_{ij}(0)=n_{ji}(0)$. In the event that the graphs for $L_p$ and $L_I$ were identical, only one set of weight updates would be necessary. The proofs and illustrations of these results are omitted in the interest of brevity, and an extended discussion will appear in a longer version of the paper.

\section{Motivating example revisited}\label{sec:example}
\begin{figure}[htpb]
\centering
  \includegraphics[width=0.5\textwidth]{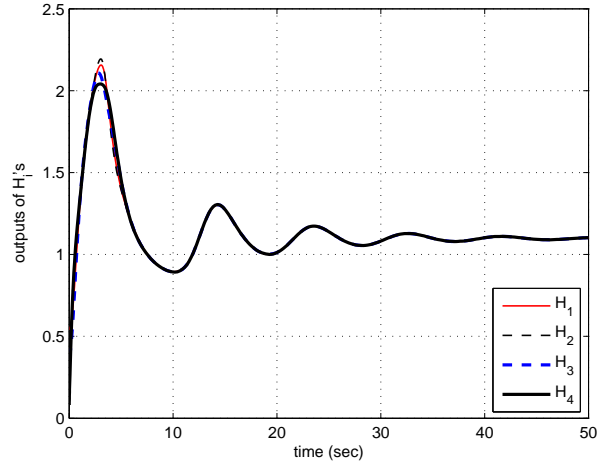}
  \caption{The outputs of the four Goodwin oscillators are synchronized with the control \eqref{fullcontrol} and \eqref{control}.}\label{gw_pi_dis}
\end{figure}

We now implement our control law, given in (\ref{fullcontrol}) and (\ref{control}), to recover the
synchronization of the outputs of the four oscillators. All nonzero $n_{ij}$ in \eqref{eqn:Li} are set to $1$. The initial conditions and the disturbance $\phi$ remains the same as in Example $2$. Fig.~\ref{gw_pi_dis} shows that the outputs of the oscillators are asymptotically synchronized.  Note
that Theorem~\ref{Thm:syn} only guarantees the synchronization of the outputs $y_i$ and may not recover the nominal oscillations of $y_i$ shown in Fig.~\ref{gw_p}. In fact, as manifested in the proof of Theorem~\ref{Thm:syn} (cf. \eqref{Vdot2} and \eqref{eq:Zdot}), the internal model based control $L_I\eta$ in \eqref{ubar} only compensates for the effects due to $\Pi\phi$, the
difference between $\phi$ and $\frac{1}{N}1_N1_N^T\phi$. Therefore, if $1_N^T\phi\neq 0$, the remaining disturbance $\frac{1}{N}1_N1_N^T\phi$ still enters the system. However, it does not affect the synchronization. 

We next present two examples where the oscillations of $y_i$ can also be recovered.

\subsubsection{$1_N^T\phi=0$} 
\begin{figure}[htpb]
\centering
  \includegraphics[width=0.5\textwidth]{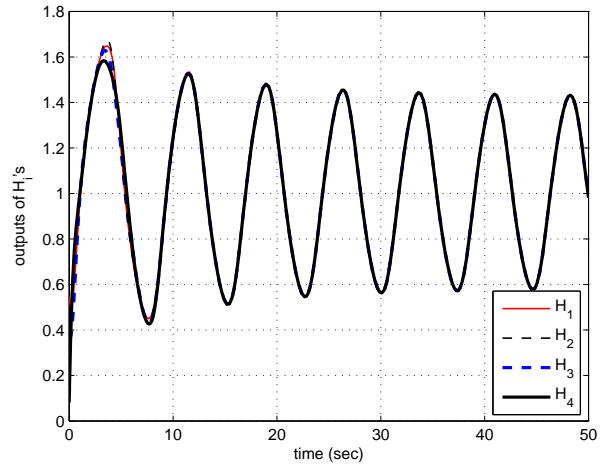}
  \caption{The outputs of the four Goodwin oscillators when $1_N^T\phi=0$. The control \eqref{fullcontrol} and (\ref{control}) recovers the synchronization of the outputs,
   which exhibit synchronized oscillations as in Example 2.}\label{gw_pi_dis0}
\end{figure}

As discussed above, if $1_N^T\phi=0$,
all the disturbances are compensated for by our control. We choose $\phi = [-0.155~0.385~-0.365~0.135]^T$ such that $1_N^T\phi = 0$. The simulation results in Fig.~\ref{gw_pi_dis0} illustrate that the outputs of the four Goodwin oscillators exhibit synchronized oscillations shown in Example 2. 
\subsubsection{Synchronize with a reference}
\begin{figure}[htpb]
\centering
  \includegraphics[width=0.5\textwidth]{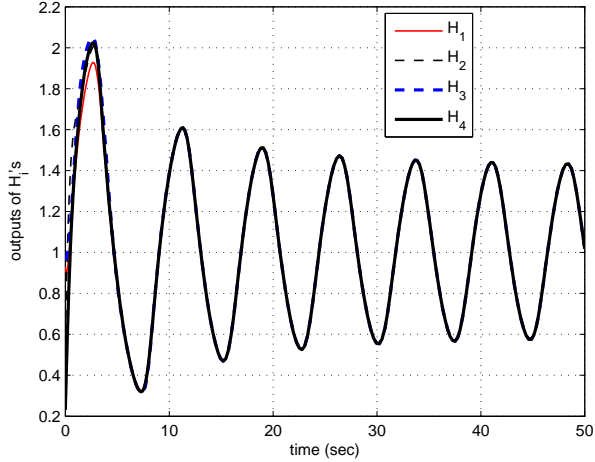}
  \caption{The outputs of the four Goodwin oscillators when $\phi_1=0$, and $\mathcal{H}_1$ employs only the proportional feedback \eqref{control:no_dis} while the other $\mathcal{H}_i$ ($i=2,3,4$) implement
  \eqref{fullcontrol} and (\ref{control}). Under this modification, the outputs also exhibit synchronized oscillations as in Example 2.}\label{gw_pi_dis_leader}
\end{figure}

In this example, we suppose that for some $\mathcal{H}_i$, say,
$i=1$, no disturbance enters $\mathcal{H}_1$, that is,
$\phi_1=0$. Then $\mathcal{H}_1$ can be considered as a reference (a
leader) and it can choose to implement $u_1=-L_p^1y$, where $L_p^1$ is the first row of $L$. The
other oscillators have the same disturbance inputs as in Example $2$ and implement \eqref{fullcontrol} and (\ref{control}). With the modification, the
simulation results in Fig.~\ref{gw_pi_dis_leader} show the recovery
of both the oscillation and the synchronization of the outputs.

\section{Design
of dynamic average consensus estimators}\label{sec:dac}
In this section, we establish the connection of the developed control in~\eqref{control} with the dynamic average consensus (DAC) estimator studied in~\cite{BaiIMP10}. For the DAC problem, the terms $\phi_i$ are considered useful inputs rather than disturbances, and the objective is to design $\mathcal{H}_i$ and $G_i$ such that the output $y_i$ asymptotically tracks the average over all $\phi_i$, i.e., 
\begin{equation}\label{DAC}
\left|y_i(t) - \frac{1}{N}1_N^T\phi(t)\right|\rightarrow 0 \textrm{ as } t \rightarrow \infty.
\end{equation}

We note that the structure shown in Fig.~\ref{nonlinearstructure} is the same as the structure of the DAC estimators studied in~\cite{BaiIMP10} (cf. \cite[Fig. 1]{BaiIMP10}). In~\cite[Theorem 2]{BaiIMP10}, three conditions were developed to ensure that the objective \eqref{DAC} is satisfied for a broad class of time-varying inputs, including constant, ramp, and sinusoidal inputs. However,~\cite{BaiIMP10} did not provide a specific design of DAC estimators that guarantees the three conditions for $\phi_i$ defined in \eqref{dis:state}-\eqref{dis:output}. We now provide a constructive approach to designing such a DAC estimator. We will show that the resulting DAC estimator consists of an IOSP $\mathcal{H}_i$ and the $G_i$ defined in \eqref{IMC:state}-\eqref{IMC:output}.  

We assume that $\mathcal{H}_i$ is a linear time invariant system. In \eqref{IMC}, we let $B_i = B$, $\forall i$. We also assume $L_p=L_I$. For the ease of discussion and comparison with~\cite{BaiIMP10}, we convert the state space representations used in Section~\ref{sec:main} to frequency domain. Towards this end, let the Laplace transforms of the disturbances $\phi_i$ in \eqref{dis:state} and \eqref{dis:output} be $\frac{c_i(s)}{d(s)}$, $i=1,\cdots,N$. Due to the skew symmetry of $A$ in \eqref{dis:state}, the $d(s)$ can be one of the following three forms:
\begin{equation}
d(s)=\left\{\begin{array}{c}
s\\
s(s^2+\omega_1^2)(s^2+\omega_2^2)\cdots(s^2+\omega_r^2),\quad
r\geq 1\\
(s^2+\omega_1^2)(s^2+\omega_2^2)\cdots(s^2+\omega_r^2),\quad
r\geq 1,
\end{array}\right.\label{dsspecial}
\end{equation}
where $\omega_i>0$ and $\omega_i\neq\omega_j$, $\forall i\neq j$. Let the order of $d(s)$ be $m$. 

We denote by $h_i(s)$ and $g(s)$ the transfer function of $\mathcal{H}_i$ from the input $u_i$ to the output $y_i$ and the transfer function of $G_i$ from $\sum_{(i,j)\in E}n_{ij}(y_i-y_j)$ to $\eta_i$, respectively. Let $n_q(s)$ and $d_q(s)$ represent the numerator
and denominator polynomials of a transfer function $q(s)$, respectively. 

We claim that the objective in \eqref{DAC} is achieved with the control in \eqref{control} and $h_i(s)$ chosen in the form of 
\begin{equation}\label{hs}
h_i(s)=h(s)=\frac{n_h(s)}{\epsilon d(s)+n_h(s)},\quad \forall i,
\end{equation}
where $n_h(s)$ is a nomic stable polynomial of order $m-1$ and $\epsilon$ is a sufficiently small positive constant. Note that when $m=1$, that is, $d(s)=s$, we choose $n_h(s)$ to be a positive constant. We prove our claim by demonstrating that the control in \eqref{control} and $h_i(s)$ in \eqref{hs} satisfy the three conditions specified in~\cite[Theorem 2]{BaiIMP10}.

 We first show that condition a) in~\cite[Theorem 2]{BaiIMP10} is satisfied. Note that $n_h(s) - d_h(s) = -\epsilon d(s)$, which ensures $d(s)|(n_h(s) - d_h(s))$. We next employ~\cite[Lemma]{highgainSPR} to show that $h(s)$ is strictly positive real (SPR) and thus $d_h(s)$ is a stable polynomial. To see this, we rewrite \eqref{hs} as
\begin{equation}
h(s)=\frac{1}{\epsilon}\frac{\frac{n_h(s)}{d(s)}}{1+\frac{1}{\epsilon}\frac{n_h(s)}{d(s)}}.
\end{equation} Because $\epsilon$ is sufficiently small and $n_h(s)$ is a stable polynomial of order $m-1$, the conditions in~\cite[Lemma]{highgainSPR} are satisfied. Therefore, $h(s)$ is SPR and $d_h(s)$ is a stable polynomial. Note that because $h(s)$ is SPR, it holds that $h(s)$ is IFOP with $\gamma > 0$.

It is easy to verify from \eqref{IMC:state} and \eqref{dsspecial} that $d_g(s)=d(s)$ and thus condition b) in~\cite[Theorem 2]{BaiIMP10} is satisfied.

Condition c) is equivalent to the stability of the transfer function $\frac{h(s)}{1+h(s)(g(s)\lambda_i^2 +\lambda_i)}$, $i = 2,\cdots,N$, where $\lambda_i>0$ is the $i$th smallest eigenvalue of $L_I$. Because $h(s)$ is SPR and $g(s)$ is passive by the construction in \eqref{IMC:state}-\eqref{IMC:output}, the negative feedback connection of $h(s)$ and $g(s)\lambda_i^2 +\lambda_i$, $\lambda_i\geq 0$, is stable~\cite{Khalil} and thus the stability of $\frac{h(s)}{1+h(s)(g(s)\lambda_i^2 +\lambda_i)}$ follows.

With the three conditions in~\cite[Theorem 2]{BaiIMP10} verified, we conclude that \eqref{DAC} is achieved.

Our choice of $h(s)$ and $g(s)$ yields a constructive passivity-based design for a DAC estimator for constant and sinusoidal $\phi_i$. Because $h(s)$ is SPR and thus IOSP, this design has the same structure shown in Fig.~\ref{nonlinearstructure}. It is a special case of~\cite[Theorem 2]{BaiIMP10}, which applies to a broader class of inputs, such as ramp signals. 

We present a simulation example below to show the effectiveness of our design of $h(s)$ and $g(s)$. 
\subsection{Simulation}
We choose $d(s)=(s^2+2^2)s$, which means that the inputs $\phi_i$
are linear combinations of a constant and a $\frac{1}{\pi}$Hz
sinusoid. We first design $g(s)$ as
\begin{eqnarray}
\dot \zeta_i&=&\left(
             \begin{array}{ccc}
               0 & 0 & 0 \\
               0 & 0 & -2 \\
               0 & 2 & 0 \\
             \end{array}
           \right)
\zeta_i+\left(
      \begin{array}{c}
        1 \\
        0 \\
        1 \\
      \end{array}
    \right)
u\label{gsstate}\\
  \eta&=&\left(
           \begin{array}{ccc}
             1& 0 & 1 \\
           \end{array}
         \right)\zeta_i.
\end{eqnarray}
Next we choose $\displaystyle h(s)=\frac{(s+0.4)^2}{\epsilon
(s^2+2^2)s+(s+0.4)^2}$. From \cite{highgainSPR}, we compute that for any
$\epsilon<1.25$, $h(s)$ is SPR. We select $\epsilon=0.01$.

\begin{figure}
\centering
  \includegraphics[width=0.5\textwidth]{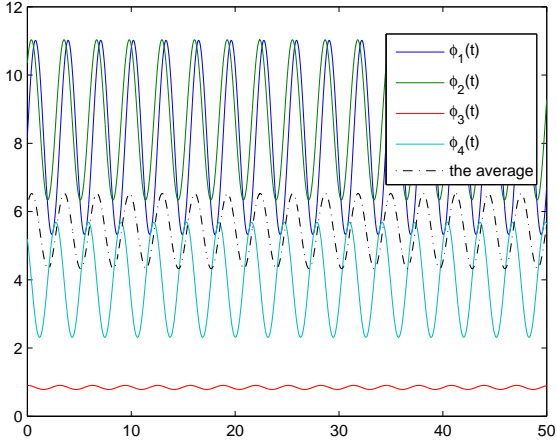}
  \caption{The input signals $\phi_i$ of the four nodes are constants plus $\frac{1}{\pi}$Hz sinusoids. }\label{inputsignal}
\end{figure}

\begin{figure}
\centering
  \includegraphics[width=0.5\textwidth]{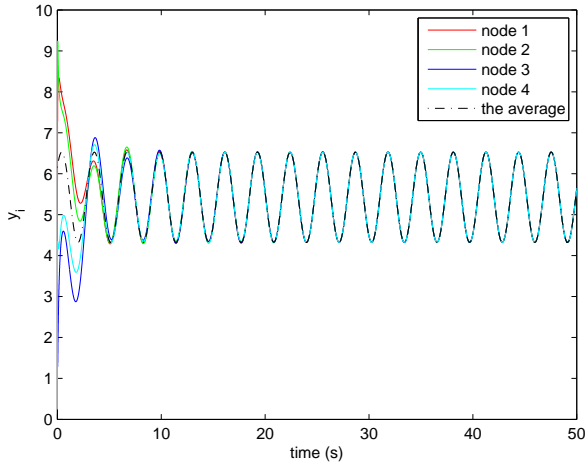}
  \caption{The output of each node converges to the average of the inputs $\phi_i$, $i=1,\cdots,4$.}\label{estimate}
\end{figure}

We consider four nodes in a cycle graph. The input signal $\phi_i$ of each node and the average of $\phi_i$ are shown in Fig.~\ref{inputsignal}. The outputs of all the nodes, $y_i$, $i=1,\cdots,4$, together with the average of $\phi_i$, are shown in Fig.~\ref{estimate}, where we observe that $y_i$ converges to the average $\frac{1}{N}1_N^T\phi(t)$. 

\section{Conclusions}\label{sec:conclusion}
We have studied synchronization of nonlinear systems that are incrementally passive, and designed a distributed control law that recovers synchronization in the presences of disturbances of a certain class using the internal model principle. Our controller has the advantage of requiring a reduced number of additional states relative to other approaches, and furthermore does not require knowledge of the initial conditions of the disturbances. The control law we proposed also provides a natural way to construct robust dynamic average consensus estimators. We have illustrated our results with several examples using Goodwin oscillators. In future work, we will demonstrate the use of adaptive updates of the coupling graph to reduce time to synchronize, and will address additional classes of disturbances and controller designs.

\bibliography{PI,mybib}
\bibliographystyle{IEEEtran}
\end{document}